\documentclass[10pt, conference, twocolumn]{IEEEtran}  
\IEEEoverridecommandlockouts                              
\usepackage{graphics,color} 
\usepackage{subfigure}
\usepackage{rotating}
\usepackage{algorithmic}
\usepackage{comment}
\usepackage{algorithm,amssymb,url}
\usepackage{epsfig}
\usepackage{flushend}
\usepackage[cmex10]{amsmath}
\usepackage{amsthm}
\usepackage{bm} 
\usepackage{cite}
\usepackage{textcomp}

\newtheorem{proposition}{Proposition}

\newcommand {\beq} {\begin{equation}}

\newcommand {\0} {\mathbf 0}

\newcommand {\bb} {{\mathbf b}}

\newcommand {\eeq} {\end{equation}}
\newcommand {\barr} {\begin{array}}
\newcommand {\earr} {\end{array}}
\newcommand {\bear} {\begin{eqnarray}}

\newcommand {\xci} {x_c^i}
\newcommand {\uci} {u_c^i}
\newcommand {\buc} {{\mathbf u}_c}

\newcommand {\nci} {N_c^i}
\newcommand {\nc}[1]{N_c^{#1}}
\newcommand {\uc}[1] {u_c^{#1}}

\newcommand {\dci} {g_c^i}
\newcommand {\dc}[1] {g_c^{#1}}
\newcommand {\Lci} {\Lambda_c^i}
\newcommand {\Lc}[1] {\Lambda_c^{#1}}
\newcommand {\C} {\mathcal C}
\newcommand {\Pci} {P_c^i}
\newcommand {\eear} {\end{eqnarray}}
\newcommand {\bears} {\begin{eqnarray*}}
\newcommand {\eears} {\end{eqnarray*}}
\newtheorem{defin}{Definition}
\newtheorem{lma}{Lemma}
\newtheorem{prop}{Proposition}

\newtheorem{thm}{Theorem}
\newtheorem{remark}{Remark}
\newtheorem{corol}{Corollary}
\begin{document}
\title{Competitive Caching of Contents \\ in 5G Edge Cloud Networks}
\author{Francesco De Pellegrini$^\diamond$, Antonio Massaro$^\diamond$, Leonardo Goratti$^\diamond$, and Rachid El-Azouzi$^\star$\thanks{$^\diamond$Fondazione Bruno Kessler, via Sommarive, 18 I-38123 Povo, 
 Trento, Italy; $^\star$CERI/LIA, University of Avignon, 339, Chemin des Meinajaries, Avignon, France. %
This research received funding from the European Union's H2020 Research and Innovation Action under
Grant Agreement No.671596 (SESAME project).}}
\maketitle
\thispagestyle{empty}

\begin{abstract}
The surge of mobile data traffic forces network operators to cope with capacity shortage. 
The deployment of small cells in 5G networks is meant to reduce latency, backhaul traffic
 and increase radio access capacity. In this context, mobile edge computing technology 
 will be used to manage dedicated cache space in the radio access network. Thus, mobile 
 network operators will be able to provision OTT content providers with new caching services 
 to enhance the quality of experience of their customers on the move.

In turn, the cache memory in the mobile edge network will become a shared resource. 
Hence, we study a competitive caching scheme where contents are stored at given price 
set by the mobile network operator.

We first formulate a resource allocation problem for a tagged content 
provider seeking to minimize the expected missed cache rate. The optimal 
caching policy is derived accounting for popularity and availability of 
contents, the spatial distribution of small cells, and the caching strategies 
of competing content providers. It is showed to induce a specific 
order on contents to be cached based on their popularity and availability.  

Next, we study a game among content providers in the form of a generalized 
Kelly mechanism with bounded strategy sets and heterogeneous players. Existence and 
uniqueness of the Nash equilibrium are proved. Finally, extensive numerical results 
validate and characterize the performance of the model.
\end{abstract}
\pagestyle{plain}


\section{Introduction}\label{sec:intro}


The recent boom of mobile data traffic is causing unprecedented stress 
over mobile networks. In fact, the global figures for such traffic 
reached 3.7 exabytes per month at the end of $2015$. They are ascribed 
mostly to over-the-top (OTT) video content providers (CP) such as Vimeo, YouTube and NetFlix. 
Forecasts predict that the world's mobile data traffic will reach $30.6$ monthly 
exabytes by $2020$, of which $75\%$ will be video~\cite{Cisco}. 

As a consequence, capacity shortage has become a real threat for mobile network operators (MNOs).  
Solutions involving the  deployment of small cell (SC) base stations~\cite{What5GwillBe} 
have been receiving large consensus from industry and academia for next 5G systems. 
SCs are low power secondary base stations with limited coverage, to which user 
equipments (UEs) in radio range can connect, hence increasing spatial reuse and network 
capacity. 

However, SCs are connected to a mobile operator's core network via backhaul technologies such
as, e.g., DSL, Ethernet or flexible millimeter-wave links. In order to avoid potential 
bottlenecks over the backhaul connection to SCs, mobile edge caching solutions have been devised.
Actually, the primary goal of mobile edge caching is precisely to circumvent the limited backhaul connection 
of SCs~\cite{WangCacheAir} and ensure fast adaptation to radio link conditions.  

From the network management standpoint, in order to handle a large number of SCs and associated memory 
caches, MNOs will rely on the emerging mobile edge computing (MEC)~\cite{ETSIMEC} 5G technology.
MEC platforms are designed to enable services to run inside the mobile Radio Access Network (RAN) 
increasing proximity to mobile users, drastically reducing round trip time and thus improving the user 
experience. 

Ultimately, CPs will be able to leverage on the  MEC caching service offered by 5G MNOs. Contents 
can be replicated directly on lightweight server facilities embedded in the radio access network 
in proximity of SCs. In this context, the design of effective mobile edge caching policies requires 
to factor in popularity, number of contents, cache memory size as well as spatial density of small 
cells to which UEs may associate to. Indeed, due to storage limitations, allocation of contents on 
mobile edge caches has become an important optimization problem~\cite{CaireFoundITW2013,Pantisano1,Poor2015,Kang2014,Sengupta2014,hachem2014coded,BastugCommMag}.

In this paper, we consider a scheme in which CPs can reserve mobile 
edge cache memory from a MNO. The MNO will provide  a multi-tenant environment 
where contents can be stored at given price and will assign the 
available caching resources to different OTT content providers. 
 In turn, this engenders competition of CPs for cache utilization.

First, we study the single CP optimization problem: under a given spatial 
distribution of SCs, the CP decides the optimal cache memory share to be reserved 
to different classes of contents. This permits to identify the minimum missed 
cache rate as a function of the purchased memory. Also, the optimal 
caching policy defines an order among contents jointly determined by two attributes: 
by the demand rate, i.e., the contents' {\em popularity}, and by the concurrent 
effect of contents with similar popularity, i.e., the contents' {\em availability}. 

Finally, the competition among CPs is formulated using a new generalized 
Kelly mechanism with bounded strategy set. CPs trade off the cost for caching contents
in the radio access network versus the expected missed cache rate. We show that the 
game admits a Nash equilibrium, and we prove that it is unique. Further properties 
of the game, including convergence and the revenue of the MNO, are investigated
numerically.

The manuscript is organized as follows. In Sec. \ref{sec:related} we provide the related work 
and we outline the main contributions. Sec.~\ref{sec:model} introduces 
the mathematical model developed throughout the paper. In Sec.~\ref{sec:contalloc} 
the optimal content caching strategy is devised and in Sec.~\ref{sec:util} we obtain 
the key characterization of the optimal missed cache rate. Sec.~\ref{sec:game} 
provides the analysis of the caching game. Numerical validation is performed in 
Sec.~\ref{sec:numres}. Finally, Sec.~\ref{sec:concls} provides closing remarks.


\section{Related Works and main contribution}\label{sec:related}


In \cite{CaireFoundITW2013} the authors consider a device-to-device (D2D) network and derive throughput 
scaling laws under cache coding and spatial reuse. Content delay is optimized in \cite{Towsley2015} by 
performing joint routing and caching, whereas in \cite{Pantisano1} a distributed matching scheme
based on the deferred acceptance algorithm provides association of users and SC base stations based 
on latency figures. 
Similarly to our model, in \cite{Poor2015} SC base stations are distributed according to a Poisson 
point process. Contents to be cached minimize a cost which depends on the expected number of missed 
cache hits. 

 In \cite{Kang2014} a model for caching contents over a D2D network is proposed. A convex optimization 
 problem is obtained and solved using a dual optimization algorithm. In our formulation 
we have obtained closed form solutions and properties of the optimal cost function.

In \cite{Sengupta2014} 
a coded caching strategy is developed to optimize contents' placement based on SC association patterns. 
 In \cite{stack} a Stackelberg game is investigated to study a caching system consisting of a content provider 
 and multiple network providers. In that model, the content providers lease their videos to the network providers to 
 gain profit and network providers aim to save the backhaul costs by caching popular videos. In \cite{hachem2014coded} the authors 
 model a wireless  content distribution system where contents are replicated at multiple access points -- depending on popularity -- so as to maximally create network-coding opportunities during delivery. Finally, \cite{BastugCommMag} proposes proactive caching in order to take advantage of contents' popularity. The scheme we develop in this work can also be applied to proactive caching.

Since content demand patterns are typically not known apriori, practical caching algorithms perform local content 
replacement policies ~\cite{Ma2004,LiPopCache}. Those rule how contents are replaced when the cache memory is full: heuristics including replacing least frequently used contents (LFU), last recently used contents (LRU) and several other variants have been proposed in literature. In our development we assume perfect knowledge of contents' popularity, namely, the demand rates: recent results \cite{LiPopCache} show that by online estimation of the contents' popularity, it is possible to achieve optimality, i.e., to minimize the missed cache rate. We leave the online estimation of the contents' demand rates as part of future works.  

{\noindent \em Main results.}  The main contributions obtained in this work are the following:
\begin{itemize}
\item a model is introduced which accounts for the contents' characteristics, 
the spatial distribution of small cells, the price for cache memory reservation and 
the effect of competing content providers under multi-tenancy;
\item using such model, by convex optimization, the optimal caching policy is found to possess a waterfilling-type of structure which induces an ordering of contents depending on contents' popularity and availability;
\item a competitive game is formulated where the price for cache memory reservation is fixed by the network provider. It is proved to be a new type of Kelly mechanism with bounded strategy set and it is showed to 
admit a unique Nash equilibrium.
\end{itemize}
To the best of the authors' knowledge, this work is the first one to study mobile 
edge caching under a competitive scheme. This appears a crucial aspect in order 
to define new business models of 5G MNOs for the emerging MEC technology. 

\begin{table}[t]\caption{Main notation used throughout the paper}
\centering
\begin{scriptsize}
\begin{tabular}{|p{0.20\columnwidth}|p{0.68\columnwidth}|}
\hline
{\it Symbol} & {\it Meaning}\\
\hline
$M$ & number of content classes\\
$\Lambda$ & intensity, i.e., spatial density of small-cells \\
$\mathcal C$ & set of content providers, $|\mathcal C|=C$ \\
$r$ & covering radius of UEs\\
$N$ & storage capacity of a local edge cache unit (number of caching slots)\\
$N_0$ & total storage capacity of the deployment \\
$\nci$ & number of contents of class $i$ for content provider $c$  \\
$\dci$ & {\em popularity}, i.e., demand rate for contents of class $i$ of content provider $c$ \\
$\Lci$ & {\em availability}, i.e., $\Lci:=\Lambda \pi r^2N/\nci$ \\
$b_c$ & caching rate of content provider $c$, $b_c\in [0,B_c]$\\
$b$ & total caching rate $b=\displaystyle  \sum_{c\in \C}b_c$ \\ 
$b_{-c}=\displaystyle  \sum_{v\not = c}b_v$ & total caching rate of competing content providers; \\                                           
$\delta$ & mobile network provider's own caching rate\\
$\buc$ & caching policy for content provider $c$, $\buc=(\uc{1},\ldots,\uc{M})$, $\sum \uci=1$\\ 
$x_c$ & share of cache memory occupied by content provider $c$\\
$\xci$ & share of cache memory for $i$-th class contents of content provider $c$\\
$B_c$ & maximum caching rate for content provider $c$\\
$\lambda_c$ & price per caching slot for content provider $c$\\
\hline
\end{tabular}\label{tab:notation}
\end{scriptsize}
\end{table}


\section{System Model}\label{sec:model}


Let us consider a MNO serving a set $\C$ of content providers, where $|\C|=C$. Each CP $c$ serves his customers leveraging the MNO network. 

Contents served to the customers of a tagged CP $c$ belong to $M$ different popularity classes, based on their demand rate or {\em popularity} $\dci$. The $i$-th popularity class thus features  $\nci$ contents and $\dci$  content requests per day. Thus, we follow a multi-level popularity model similar to the one proposed in \cite{hachem2014coded,hachem2014isit}. In such model, files  are  divided  into  different popularity  classes, and  files  within  each  class  are  equally  popular.  

\begin{figure}[t]
\centering
\includegraphics[width=\linewidth]{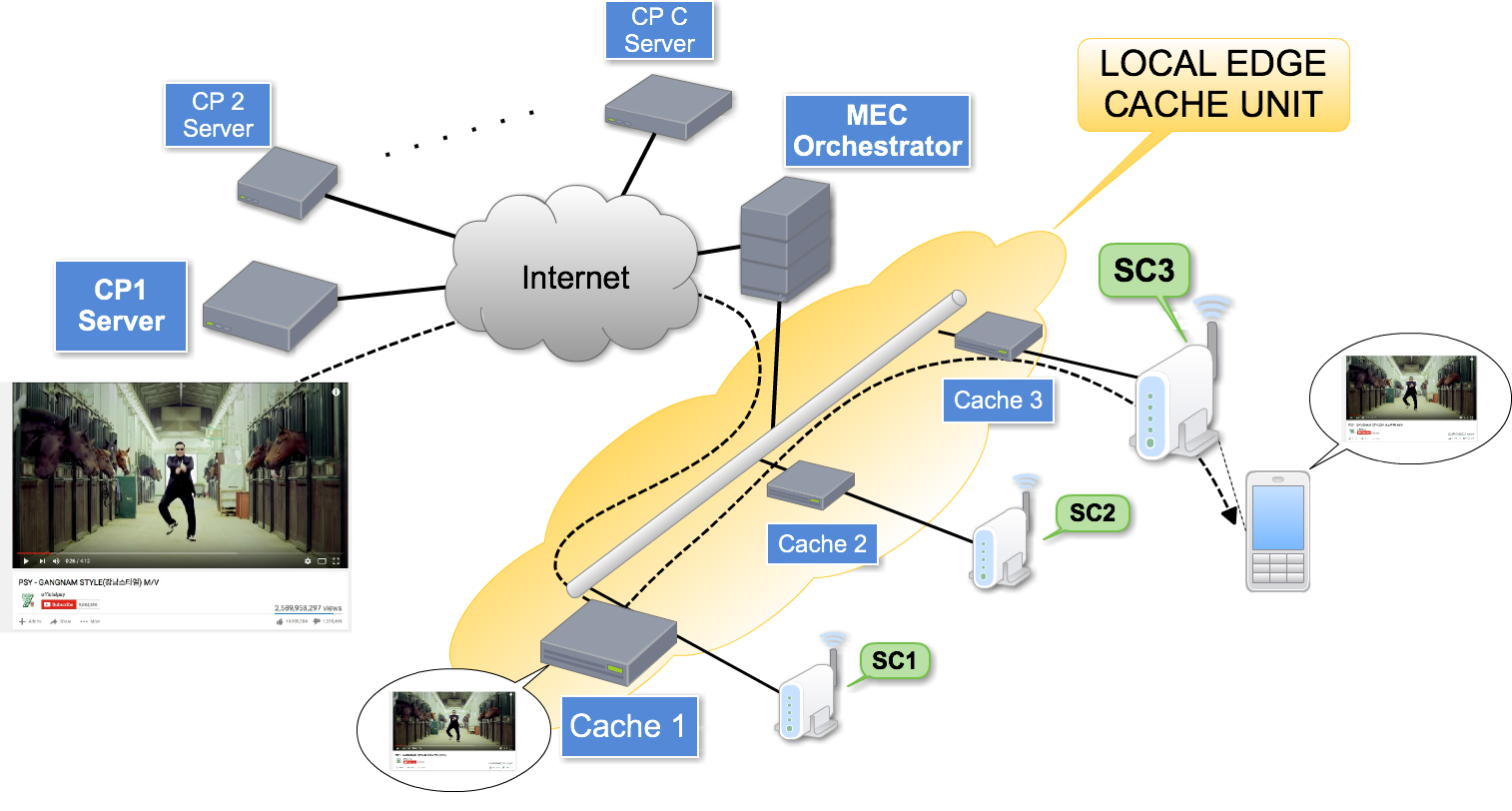}
\caption{Local edge cache unit providing $N$ memory slots.}\label{fig:edgecloud}
\end{figure}

We assume that each SC is attached to a local edge caching server, briefly {\em cache}. Multiple 
caches are aggregated by connecting them through the MNO backhaul and managed using a local MEC orchestrator, thus forming a seamless local edge cache unit as in Fig.~\ref{fig:edgecloud}. $N$ {\em caching slots} represent the available memory on such local edge cache unit; the total cache space across the whole deployment is hence $N_0=K\cdot N$ where $K$ is the number of local edge cache units. For the sake of simplicity, each content is assumed to occupy one caching slot; since we assume $N_0,N\gg 1$, we rely on fluid approximations to describe the dynamics of cache occupation. 

Fetching a non cached content from the remote CP server beyond the backhaul comes at unitary 
cost; such cost may represent the content's access delay or the throughput to 
fetch the content from the remote server. Conversely, such cost is negligible if the user associates 
to a small cell storing a cached copy of the content. However, such cache should be reached by 
connecting to a SC within the UE radio range $r>0$. SCs are distributed according to a spatial 
Poisson point process with intensity $\Lambda$. 

The following assumptions characterize the caching process:\\
\noindent i. each CP $c$ can purchase edge-caching service from the MNO and issue $b_c$ 
caching slot requests per day; we call $b_c$ the {\em caching rate}, where $0 \leq b_c \leq B_c$;\\
\noindent ii. MNO will reserve $\delta>0$ caching slots per day for her own purposes; \\
\noindent iii.  reserved slots expire after $1/\eta$ days for $\eta>0$;\\
\noindent iv. in order to attain $b_c$ caching slots per day, CP $c$ bids $\tilde b \in [0,1]$, and the MNO 
grants $b_c = b_0 \, \tilde b$ caching slots per day, where $b_0$ is such that $\sum b_c + \delta \leq N_0$. 
In our analysis we assume $b_0=1$ for the sake of simplicity\footnote{We refer to \cite{basarolsder} 
for an in depth discussion of the connection between mechanisms and fair share of resources of the type 
studied in this paper.}.\\ 
\noindent v. CPs are charged based on the caching rate $b_c$;\\
\noindent vi. demand rates $\dci$ per content class are uniform across the MNO's network. 

The MNO will thus accommodate $X_c$ memory slots for CP $c$ according to 
\begin{equation}\label{eq:state}
\dot X_c=b_c -\eta \, X_c, 
\end{equation} 
so that the whole cache memory occupation will be ruled by 
\begin{equation}\label{eq:stateall}
\dot X=b -\eta \, X, 
\end{equation} 
where $b:=\sum_c b_c + \delta$ is the total caching rate. The corresponding 
dynamics for the fraction of reserved cache memory, assuming $X(0)=0$ is 
\[
x(t)=\min \left  \{ 1, \frac{b}{N_0 \eta}\left(1-e^{-\eta t}\right) \right \}
\]
The MNO, in order to ensure full memory utilization, will choose $\eta$ such that 
$b/(N_0\eta)\geq 1$. It follows from a simple calculation that in steady state, 
the fraction of the caching space for content provider $c$ is 
\begin{equation}
x_c(t)=\frac{b_c}{b_c+b_{-c}+\delta} 
\end{equation}
Because contents' requests are uniform across the MNO's network, same fraction of cache space is occupied 
by CP $c$ in each local edge cache unit. 

In particular, CP $c$ will split his reserved memory among content classes according to a proportional share allocation with weighting coefficients $\uci$, $i=1,\ldots,M$, where $\sum_{i=1}^M \uci=1$. We define $\buc:=(\uc{1},\ldots,\uc{M})$ 
the {\em caching policy} of CP $c$. 

Then, the fraction of local edge cache memory occupied by contents of class $i$ from content provider $c$ is
\begin{equation}\label{eq:share}
\xci=\frac{ b_c}{\sum_{v \in \C} b_v +\delta}\uci
\end{equation} 

Finally, a tagged content of class $i$ of content provider $c$ is found in the memory of a local edge cache  
with probability $\Pci=\min\{\frac{N}{\nci}\xci,1\}$. In the rest of the paper, we will assume 
$N<\nci$ for the sake of simplicity.

Now, we want to quantify the probability for a given requested content not to be found in the local edge 
cache memory, i.e., the \emph{missed cache} probability. 

 Under the Poisson assumption, the probability for a tagged UE not to find any SC within a distance $r$ is $e^{-\pi r^2 \Lambda}$. Applying a thinning argument, the probability not to find a content of class $i$ of CP $c$ within distance $r$ is $e^{-\pi r^2 \Lambda P_c^i}$. 

The expected missed cache rate (MCR) is thus 
\begin{equation}\label{eq:costfun}
U_c(b_c,b_{-c},\buc)=\sum_i \dci \, e^{-\pi r^2\Lambda \frac{N}{\nci}\frac{b_c}{b_c+b_{-c}+\delta}\uci }
\end{equation}
It depends on caching rate $b_c$ and on caching policy $\buc$. Also,
 $b_{-c}:=\sum_{v\not = c}b_v$ accounts for the fact that other content providers 
share the same cache space. 
In the next section we shall describe the optimal caching policy $\buc^*$ attained when CP $c$ aims at minimizing \eqref{eq:costfun}, for a fixed value $b_c$ of the caching rate. 
 
\section{Optimal Caching Policy}\label{sec:contalloc}


In order to analyse the model introduced before, we need to characterize the 
CPs' response to competitors' actions, i.e., $b_{-c}$. Hence, we assume that each CP 
aims at minimizing his own MCR, and that the network provider guarantees full information 
to CPs, i.e., storage capacity, spatial density of SCs and total caching rate. We defer the study 
of the system under partial information at the content provider's side to later works.

We hence consider the following resources allocation problem for the single CP: 
\begin{defin}[Optimal Caching Policy]
Given opponents' strategy profile ${\mathbf b}_{-c}=(b_1,\ldots,b_{c-1},b_{c+1},\ldots,b_C)$ the optimal 
caching policy of $c\ \in \C$ is the solution of 
\begin{equation}\label{eq:best_resp}
\mathbf{u}_c^* := \arg\min_{\uc{1},\ldots,\uc{M}} U_c(b_c,b_{-c},\buc)
\end{equation}
subject to the following constraints:
\begin{equation}\label{eq:constraints}
\uci\geq0, \quad \sum_i\uci = 1
\end{equation} 
\end{defin}
It is immediate to observe that $U_c(b_c,b_{-c},\buc)$ is a strictly convex function in the single content provider control 
$\buc$, so that a unique solution exists \cite{boyd}. In order to solve for the constrained 
minimization problem in equations \eqref{eq:best_resp} and  \eqref{eq:constraints} we can write the Lagrangian 
for player $c \in \C$ as  follows
\begin{equation}
L_c(\mathbf{u}_c,\mathbf{\mu},\nu)=\sum_i \dci e^{-\Lci\frac{b_c}{b+\delta} \uci}%
-\sum_i\mu_i\uci+\nu\left(\sum_i\uci-1\right)\nonumber
\end{equation}
For notation's sake, we have defined $\Lci := \pi r^2\, \Lambda_f \frac{N}{N^c_i}$; we define this 
quantity {\em availability}. Furthermore, since constraints are affine, the Karush Kuhn Tucker (KKT) 
conditions provide the solution of the original problem \cite{boyd}.

Hereafter, we enlist the KKT conditions: 
\begin{eqnarray*}
&& \nabla_u L_c(\mathbf{u}_c,\mathbf{\mu},\nu)={\mathbf 0}
\qquad  \mbox{stationarity}\\ 
&&\uci\geq 0  \qquad \mbox{primal feasibility: control}\\
&&\sum_i\uci-1= 0  \qquad \mbox{ primal feasibility: normalization}\\
&&\mu_i\geq 0  \qquad \mbox{ dual feasibility: control}\\
&&\nu\geq 0  \qquad \mbox{ dual feasibility: normalization}\\
&&\mu_i\uci=0  \qquad \mbox{complementarity slackness}
\end{eqnarray*}

Using a standard argument \cite{boyd}, by complementary slackness, $\uci>0$ implies 
$\mu_i=0$; let us define index set $I:=\{i \in \{1,2,\ldots,M\}\,|\, \uci>0\}$.


\subsection{Popularity sorted case}\label{sec:popsort}

Let us discuss a simplified setting where popularity is the main driver for the CPs. 
Let us first assume that the indexes are sorted according to contents' popularity $\dc{r}$, 
i.e., $\dc{1}\geq \ldots \geq \dc{M}$. We also assume $\nc{1} \leq \ldots\leq \nc{M}$: 
more popular contents are also less abundant. This assumption will be relaxed in the next 
section, where we derive the general solution; we will see that there exists a natural 
order combining popularity and availability of contents which determines whether a
 content class is cached or not.

In the case at hand, the structure of the optimal allocation follows from
\begin{lma}\label{lemma:saturation}
Let $\dc{r}\geq  \dc{r+1}$ and $\nc{r}\leq \nc{r+1}$ for $r=1,\ldots,M$, then ${u^*}_c^r=0$ 
implies ${u^*}_c^{r+1}=0$ 
\end{lma}

From the previous statement we can deduce the following 
\begin{corol}\label{eq:corl1}
Under the assumptions of Lemma.~\ref{lemma:saturation} there exists 
$1\leq r_0 \leq M$ such that response ${u^*}_c^i>0$ for $i \leq r_0$ and 
${u^*}_c^i=0$ otherwise
\end{corol}
The stationarity conditions can be used in order to determine the optimal content 
allocation in closed form. Let $0\leq i \leq r_0$, then $\mu_i=\mu_{r_0}=0$, so that 
\[
\Lc{r_0}\dc{r_0} e^{-\Lc{r_0}\frac{b_c}{b+\delta} {u^*}_c^{r_0}}=\Lci\dci e^{-\Lci\frac{b_c}{b+\delta}{u^*}_c^{i}}
\]
and 
\[
{{u^*}_c^i}=\frac{\Lc{r_0}}{\Lc{i}}{u^*}_c^{r_0} - \frac{b+\delta}{ \Lci b_c} \log \Big ( \frac{\dc{r_0}\,\Lc{r_0}}{\dc{i}\,\Lc{i}}\Big ) 
\]

Finally, due to the constraint saturation 
\begin{eqnarray}\label{eq:threshold}
{u^*}_c^{r_0} = \frac{1 + \frac{b+\delta}{b_c}\sum_{i=0}^{r_0} \frac{1}{\Lc{i}}\log \Big ( \frac{\dc{r_0}\,\Lc{r_0}}{\dc{i}\,\Lc{i}}\Big )}{\sum_{i=0}^{r_0}\frac{\Lc{r_0}}{\Lc{i}}} 
\end{eqnarray}
From Corol.~\ref{eq:corl1}, the optimal solution corresponds to the maximal $r_0$ such that the corresponding 
${u^*}_c^{r_0}$ solving \eqref{eq:threshold} lies in $[0,1]$. 

We hence observe that when $\Lci=\Lc{j}$ for all $i,j$, i.e., when availability is same for all classes, the optimal caching policy depends on contents' popularity only. The smaller the request rate $\dci$, i.e., the less popular the content class, the smaller the cache share reserved to contents of that class. Such optimal policy is analogous to the optimal content replacement $\texttt{MIN}$ algorithm~\cite{BeladyMIN}. In fact, $\texttt{MIN}$, 
assumes full information about the future, replaces first contents which will be requested farthest in the future. 

\subsection{General solution}

The solution to the KKT conditions can be formulated as a waterfilling-like solution \cite{boyd}. 
In fact, from stationarity conditions, $\mu_r$ writes as 
\[
\partial_{u_c^i}L_c=-\Lci\frac{b_c}{b+\delta}\dci e^{-\Lci\frac{b_c}{b+\delta} \uci}-\mu_i+\nu=0
\]
\[
\mu_i=\nu-\Lci\frac{b_c}{b+\delta}\dci e^{-\Lci\frac{b_c}{b+\delta} \uci}
\]

which can be specialized into the following two cases. 

\noindent {\em Case i}: $\nu>\Lci\dci\frac{b_c}{b+\delta}$. In this case $\mu_i>0$ for any $\uci\geq 0$. 
Hence, by complementary slackness, $\uci=0$.

\noindent {\em Case ii}: $\nu\leq\Lci\dci\frac{b_c}{b+\delta}$. It is always possible to find $\uci>0$ 
satisfying the stationarity condition and a $\mu_i$ that satisfies the complementary slackness condition: just set $\mu_i=0$ and
\[
\uci=\frac{b+\delta}{\Lci b_c} \log\left(\frac{\Lci\dci}{\nu}\frac{b_c }{b+\delta}\right)
\]
Finally, let $\alpha_i:=\frac{b+\delta}{\Lci\dci b_c}$. For notation's sake, the solution writes
\begin{eqnarray}\label{eq:watersol1}
&&\hskip-12mm {u_c^*}^i=\begin{cases}
\frac{b+\delta}{\Lci b_c}\left(\log(1/\nu) -\log(\alpha_i)\right)& \mbox{ if } 1/\nu>\alpha_i\\
0 & \mbox{ if } 1/\nu \leq \alpha_i\\
\end{cases}\\
&&\hskip-12mm \mbox{subject to: } \quad \sum_i {u_c^*}^i =1\nonumber
\end{eqnarray}

It is immediate to recognize a waterfilling solution in logarithmic scale.
Let $\alpha=\min_i\alpha_i$. Indeed $\sum_i{u_c^*}^i $ is strictly increasing in $1/\nu, 1/\nu>\alpha$.  
Also, $\sum_i{u_c^*}^i(1/\nu)=0$ for $1/\nu\leq \alpha$, and $\lim_{1/\nu\rightarrow\infty}\sum_i{u_c^*}^i(1/\nu)=\infty$. Thus, there exists a unique positive $\nu$ satisfying our problem.  

Actually, the solution is determined in polynomial time $O(M)$: let $[\cdot]$ be the permutation of the 
indexes which sorts $\alpha_i$ in increasing order, i.e., $\alpha_{[i]}\leq \alpha_{[i+1]}$. For every choice 
$\alpha_{[i]}\leq 1/\nu  \leq \alpha_{[i+1]}$, one can determine a value of $\nu$ 
\[
\log(1/\nu)=\frac{\frac{b_c}{b_c+b_{-c}+\delta}+\sum_{r=1}^k \frac{\log \alpha_{[r]}}{\Lc{[r]}}}{\sum_{r=1}^k \frac 1{\Lc{[r]}}}
\]
for $k=1,\ldots,M$. Then, consider the only $1/\nu$, compatible with \eqref{eq:watersol1}. We observe that $\alpha_{[i]}\leq \alpha_{[i+1]}$ is equivalent to state that $\dc{[i]}\Lc{[i]} \geq \dc{[i+1]}\Lc{[i+1]}$: clearly, if $u_c^{[i]}=0$, indeed $u_c^{[i+1]}=0$, so that we can generalize Prop.~\ref{eq:corl1} as follows
\begin{corol}[Threshold structure]\label{eq:corl2}
There exists $1\leq r_0 \leq M$ such that  ${u_c^*}^{[s]}>0$ for $s \leq r_0$ and ${u_c^*}^{[s]}=0$ otherwise.
\end{corol}
\begin{remark}[Contents' Order]\label{rem:popularity}
The existence of a threshold structure in a waterfilling-type of solution is not surprising; what we learn instead is that the natural order 
which determines which content classes are cached or not is given by the values $\dci\cdot \Lci$. Hence, the index sorting $[\cdot]$ which 
orders the content classes with decreasing $\dci\cdot \Lci$ is the order by which a content provider prioritizes content classes to be cached as the cache memory available increases.
\end{remark}
In the rest of the paper we assume content classes sorted according to $[\cdot]$. 


\section{Optimal Missed Cache Rate}\label{sec:util}


CPs who optimize contents to be cached, for a given value of 
$b_c$, minimize the expected MCR $U_c(b_c,b_{-c},\buc)$ in the caching policy $\buc$. 
In the game model presented in the next section we shall leverage on the convexity
properties of the optimized MCR $U_c: \mathbb {R}_+^2 \rightarrow \mathbb {R}$, defined as 
\begin{equation}\label{eq:onedymutil}
\hskip-3mm U(b_c,b_{-c}):=\min_{\mathbf \buc \in \Pi}\!\! %
\left \{ \sum_i \dci e^{-\Lci\frac{ b_c}{b_c + b_{-c}+\delta}\uci} \right \}
\end{equation}
where $\Pi=\{\mathbf u \in \mathbb{R}^M| \buc \geq 0, \sum \uci = 1\}$. As already proved, the 
minimum in \eqref{eq:onedymutil} is unique, hence $U_c(b_c,b_{-c})$ is well defined. Hereafter 
we thus demonstrate its convexity in $b_c$. 

Actually, convexity can be derived for a class of functions 
wider than the posynomial expression appearing in \eqref{eq:onedymutil}. We first need  the following 
fact, whose proof is found in the Appendix.
\begin{lma}\label{lemma:lemma3}
Let $f$ be non increasing, with domain $\mathbb{R}_+$. Let $H(x)=x \, f(x)$ be convex. 
Then $f$ is convex on $\mathbb{R}_+$. 
\end{lma}

We can now derive the general conditions for the convexity of the optimal missed cache rate 
\begin{thm}\label{lem:conv}
Let $h: \mathbb{R}^{M} \rightarrow \mathbb{R}$, convex and decreasing in each variable $x_i$ for
$i=1,\ldots,M$, then 
\[
U_c(b_c,b_{-c}):= \min_{\mathbf \buc \in \Pi} h \Big (\frac{\uc{1} b_c}{b_c + b_{-c} + \delta},\ldots, \frac{\uc{M} b_c}{b_c + b_{-c} + \delta}\Big )  
\]
is convex and decreasing in $b_c$.
\end{thm}
\begin{IEEEproof}
In order to prove convexity for $U_c$, we consider perspective function $H(t,\mathbf x)=t \cdot h(\mathbf x/t)$: $H$ is known to be convex if $h$ is convex \cite[pp.89]{boyd}. In the next step, let $a=b_{-c}+\delta$, and consider the function 
\begin{eqnarray} 
\hskip-7mm &&\min_{\mathbf (\uc{1},\ldots,\uc{M})} \left \{ H(b_c + a,b\cdot \buc)\left | \sum \uci=1 ,\uci\geq 0\right .\right \} \nonumber  \\
\hskip-7mm &&=(b_c + a) \cdot  \!\!\!\!\min_{\mathbf (\uc{1},\ldots,\uc{M})} \left \{ h\Big (\frac{b_c \cdot \buc}{b_c + a} \Big )\left | \sum \uci=1,\uci\geq 0 \right .\right \}\nonumber \\
\hskip-7mm &&=(b_c + a) \,  U_c((b_c + a)-a,b_{-c})=(b_c + a) \, \widehat U_c(b_c + a ,b_{-c}) \nonumber 
\end{eqnarray}
which is convex since it is obtained by minimizing $H(b_c + a,\mathbf x)$ over the simplex $\sum x_i=b_c$ which is a convex set. 
%
Now using Lemma~\ref{lem:conv}, we conclude that $\widehat U_c(b_c + a,b_{-c})$ is convex in the first variable, and by affinity so does $U_c(b_c,b_{-c})$. 

In order to prove the monotonicity of $U_c(b_c,b_{-c})$ in $b_c$, let us consider $b_c\geq 0$ and $b_c+\epsilon$ for some $\epsilon>0$ and the respective optimal caching policy $\buc^*(b_c)$ and $\buc^*(b_c+\epsilon)$. We write

\begin{small}
\begin{eqnarray}
\hskip-3mm&&h\Big (\buc^*(b_c) \frac{b_c}{b_c+b_{-c}+\delta} \Big ) > h\Big (\buc^*(b_c) \frac{b_c+\epsilon}{b_c+\epsilon+b_{-c}+\delta} \Big ) \nonumber\\
&&> h\Big (\buc^*(b_c+\epsilon) \frac{b_c+\epsilon}{b_c+\epsilon+b_{-c}+\delta} \Big )\nonumber 
\end{eqnarray}
\end{small}
where the first inequality follows from monotonicity and the second from optimality.
\end{IEEEproof}

The case in \eqref{eq:onedymutil} satisfies the assumptions by letting $h(\mathbf x)=\sum_i \dci e^{-\Lci\, x_i}$. 

For presentation's sake, in Sec.~\ref{sec:game} we shall identify $U(b_c,b_{-c},\buc^*):=U(b_c,b_{-c})$. There, we also need the following result, whose proof is found in the Appendix. 
\begin{lma}[Limit solution for $b_c\rightarrow 0$]\label{lma:limit} 
There exists $\varepsilon>0$ such that, for any $b_c<\varepsilon$, $\buc^*=(1,0,\ldots,0)$ and the optimal MCR is
\begin{equation}\label{eq:utilzero}
U_c(b_c,b_{-c},\buc^*)=\dc{1} e^{-\Lc{1}\frac{b_c}{b_c + b_{-c}+\delta}}+\sum_{i > 1}\dc{i}
\end{equation}
\end{lma}

\subsection{The case $M=2$} 

For two classes of contents, $M=2$, the expression for $U_c(b_c,b_{-c})$ can be derived in simple closed form. 
This sample case retains the main properties of the optimal policy and it is useful in order to provide insight into the structure of the optimal MCR. First, we write the expression of the optimal MCR
\begin{equation}
U_c(b_c,b_{-c})=\min_{0\leq \uc1 \leq b_c} \dc{1} e^{-\Lc{1} \frac{b_c\,\uc1}{b_c+b_{-c}+\delta}} +\dc{2} e^{-\Lc{2} \frac{b_c(1-\uc1)}{b_c+b_{-c}+\delta}}
\end{equation}
For the sake of notation, we denote $\Gamma:=\frac{\dc{2}\Lc{2}}{\dc{1} \Lc{1}}$. The (unconstrained) minimum of the right hand term is attained at 
\begin{equation}\label{eq:min2D}
{u^*}_c^1=\frac{\Lc{2}}{\Lc{1}+\Lc{2}}-\frac{(b_c+b_{-c}+\delta)}{b_c(\Lc{1}+\Lc{2})}\log(\Gamma)
\end{equation}
When ${u^*}_c^1\in (0,1)$, the utility function of $c\in \C$ is 
\[
U_c(b_c,b_{-c})= K_c \, e^{-\frac{\Lc{1}\Lc{2}}{\Lc{1}+\Lc{2}}\frac{b_c}{b_c+b_{-c}+\delta}}
\]
where the constant appearing on the first term is 
\begin{equation}\label{eq:Kc}
K_c=\dc{1} \cdot  \Gamma ^{\frac{\Lc{1}}{\Lc{1}+\Lc{2}}}+ \dc{2} \cdot \Gamma ^{-\frac{\Lc{2}}{\Lc{1}+\Lc{2}}}  
\end{equation}
Incidentally, the convexity of $U_c(\cdot,b_{-c})$ for $M=2$ can be verified directly from the convexity of $\exp(1/x)$ and by composition with an affine function, which preserves convexity. 

We are interested in characterizing precisely the behavior of the expected MCR as a function 
of $b_c$. In particular, we want to assess the influence of the system parameters.


Now, we can obtain the following result
\begin{proposition}\label{prop:2Dcase}
i. Assume $\Gamma<1$. Let $\Lc{1} > \log(1/\Gamma)$, and define threshold
 for content $2$
\begin{equation}\label{eq:bcstar}
b_c^{\star}=(b_{-c}+\delta)\frac{\log(1/\Gamma)}{\Lc{1} - \log(1/\Gamma)}
\end{equation}
then it holds
\begin{equation}\label{eq:util}
U_c(b_c,b_{-c})=\begin{cases}
\dc{1} e^{-\Lc{1} \frac{b_c}{b_c+b_{-c}+\delta}} + \dc{2} & \mbox{if} \quad 0 \leq b_c <  b_c^{\star}\\
K_c \, e^{-\frac{\Lc{1}\Lc{2}}{\Lc{1}+\Lc{2}}\frac{b_c}{b_c+b_{-c}+\delta}} & \mbox{if} \quad b_c \geq b_c^{\star}
\end{cases}
\end{equation}
where the corresponding optimal cache policy is $(1,0)$ in the first case,  $({u^*}_c^1,1-{u^*}_c^1)$ in the 
second case and constant $K_c$ as in \eqref{eq:Kc}\\
ii. Let $\Lc{1} \leq  \log(1/\Gamma)$, then $(1,0)$ case holds for any $b_c>0$ with associated expected MCR
defined as in case i. \\
iii. If $\Gamma>1$, both i. and ii. hold with role of content $1$ and $2$ reversed.
\end{proposition}
The proof follows by inspection of \eqref{eq:min2D} considering ${u^*}_c^1$ as an unconstrained 
minimizer. First, we observe that if $\Gamma<1$, then ${u^*}_c^1>0$, i.e., the first content class is always cached. The other conditions follow by imposing ${u^*}_c^1 \geq  1$. 

\subsection*{Discussion: availability, popularity and competition} 

Hereafter we draw insight from Prop.~\ref{prop:2Dcase}. First, as seen 
there, the optimal caching rate ${u^*}_c^1$ depends solely on a few 
system parameters, namely $\dc{i}$ and $\Lc{i}$ for $i=1,2$. Actually, when 
$\Gamma<1$ then ${\uc{1}}^*=1-{\uc{2}}^*>0$: contents of type $1$ are always 
cached because $\dc{2}\Lc{2}<\dc{1}\Lc{1}$. The fact that contents of 
type $2$ are cached depends on the sign of $\Lc{1} - \log(1/\Gamma)$, which
in turn determines the actual structure of the waterfilling solution.

From Prop.~\ref{prop:2Dcase}, $\Gamma<1$ means that contents of type $2$ are either 
less popular ($\dc{2}\leq \dc{1}$) and/or less available ($\Lc{2}\leq \Lc{1}$) than 
contents of type $1$. The availability $\Lc{1}$ of contents of type $1$ determines whether 
they will be eventually cached. In practice, when $\Lc{1} > \log(1/\Gamma)$, there exists 
a critical value of the CP caching rate $b_c$, i.e., the threshold \eqref{eq:bcstar}. 
Above such value, contents of type $2$ are cached, below that they are not cached. For 
the sake of consistency, in the case when $\Lc{1} \leq \log(1/\Gamma)$, $b^\star=+\infty$ 
while for $\Gamma>1$, $b^\star=0$.

Furthermore, $b_c^\star$ increases linearly with both the MNO caching 
rate $\delta$ and the competitors' aggregate caching rate $b_{-c}$: competition 
for edge caching resources tends to prevent caching of contents with smaller product $\dci \Lci$. Actually, 
under higher competition figures, optimal caching policies are of the type ${\uc{1}}^*=1$, 
and ${u^*}_c^2=0$. It is interesting to observe that, as detailed in case ii., not always there 
 exists a caching rate $b_c$ such that it is worth caching the least profitable content 
 class.  

\begin{figure}[t]
    \centering
   	  \subfigure
    {
        \includegraphics[width=0.46\linewidth]{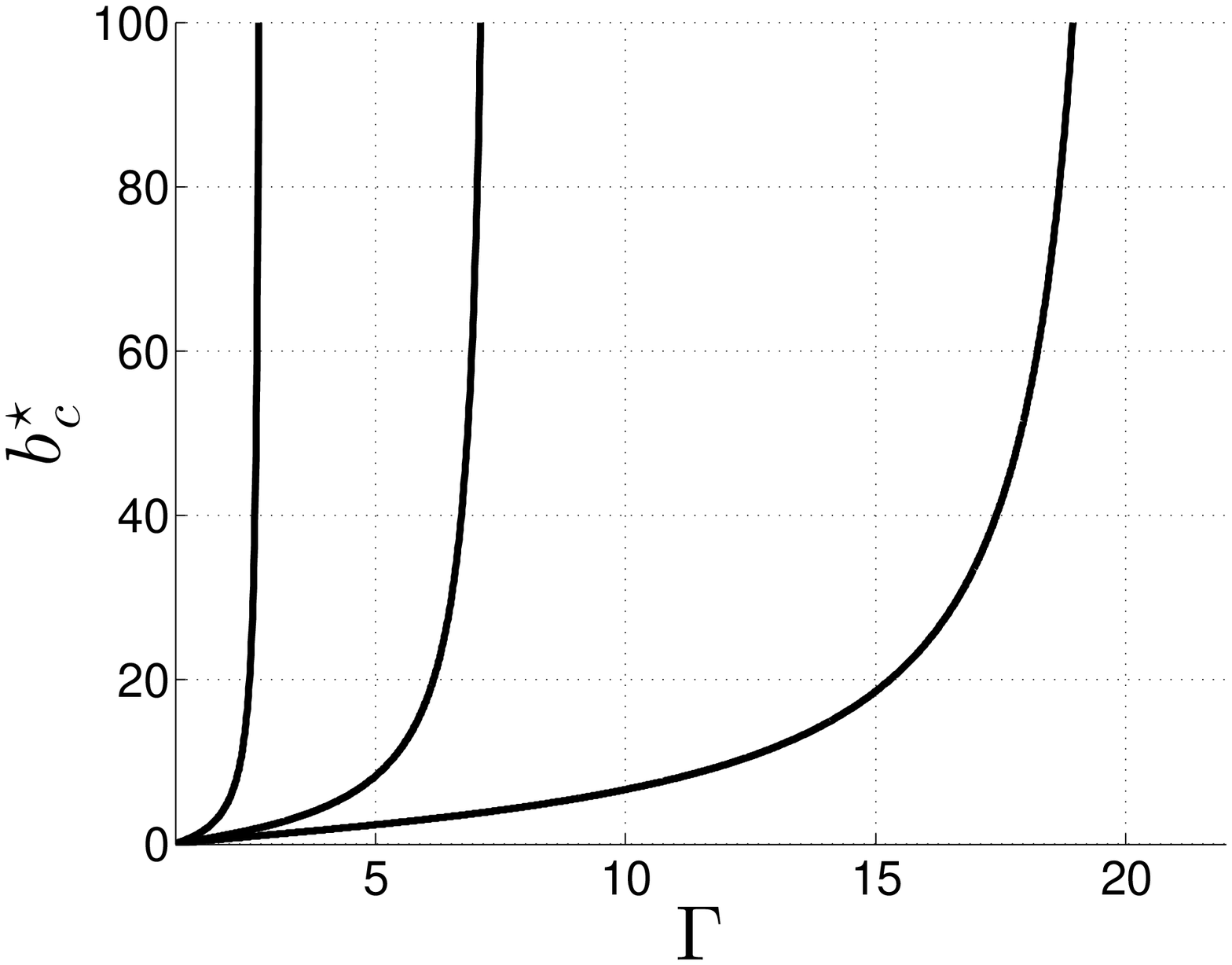}
        \put(-120,85){$(a)$}
        \put(-90,55){\begin{turn}{90}{$\Lc{1}=1$}\end{turn}}
        \put(-69,55){\begin{turn}{90}{$\Lc{1}=2$}\end{turn}}
        \put(-15,55){\begin{turn}{90}{$\Lc{1}=3$}\end{turn}}
        \put(-60.5,0){\small $1\!/$}
        \label{fig:2Da}
    }
        \subfigure
    {
        \includegraphics[width=0.46\linewidth]{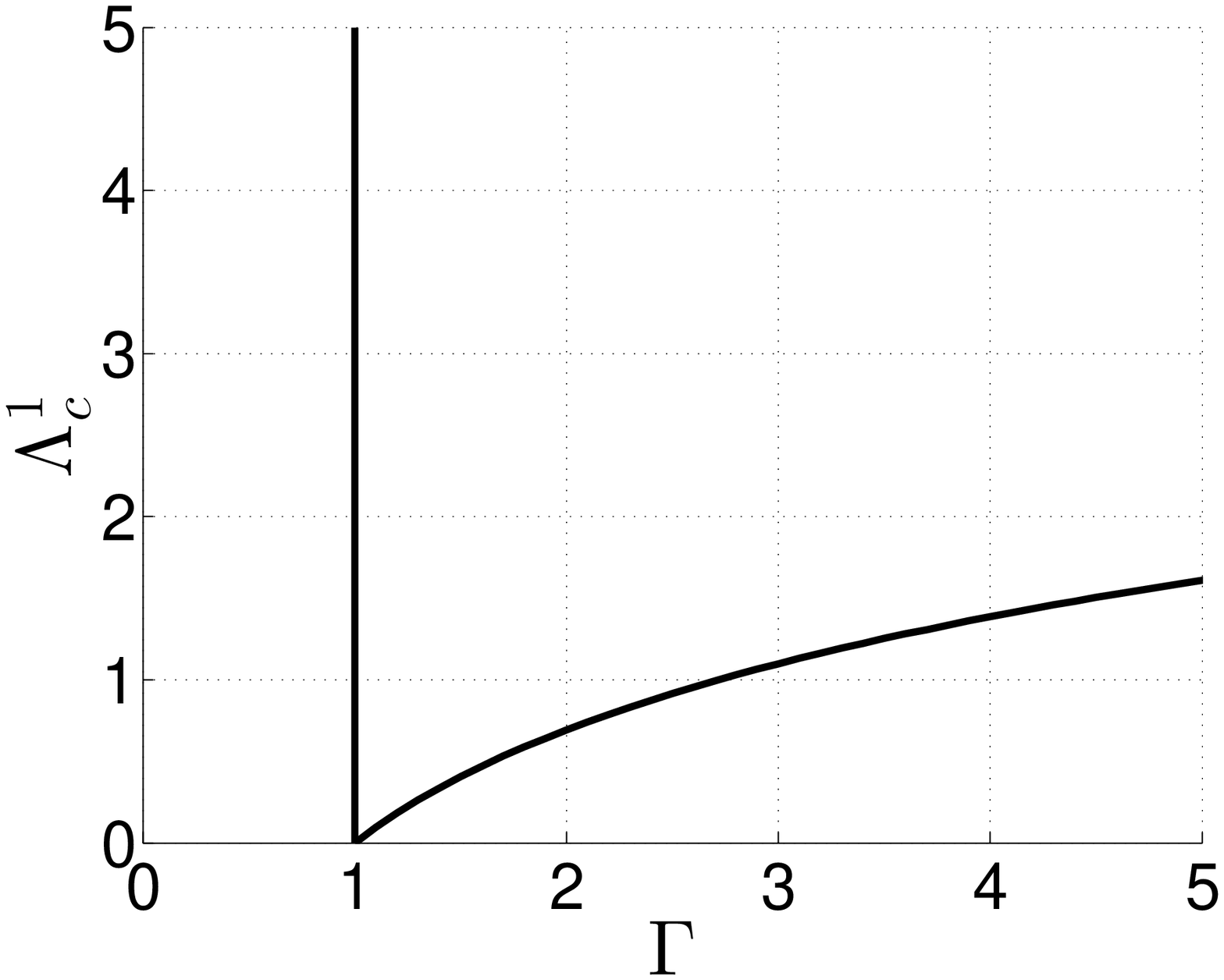}
        \put(-120,85){$(b)$}   
        \put(-94,60){\begin{turn}{90}{$b^\star=0$}\end{turn}}
        \put(-70,60){$b^\star<\infty$}
        \put(-40,20){$b^\star=\infty$}
        \put(-63.5,0){\small $1\!/$}
        \label{fig:2Db}
    }           
       \caption{Case $M=2$: (a) Increasing value of $b_c^\star$ as a function of $\Gamma$, %
       for $\Lc{1}=1,2,3$ and $u_{-c}=\delta=1$ (b) Region of switch on of content $2$.}\label{fig:2D}
\end{figure}

We have provided a pictorial representation of the results of this section in Fig.~\ref{fig:2D} for 
the case $M=2$. In Fig.~\ref{fig:2Da} the value of $\Lc{1}$ has been fixed at different values and 
the corresponding behavior of the threshold value $b_c^\star$ has been reported as a 
function of $\Gamma$. For $\exp(\Lc1) \leq 1/\Gamma$, it holds $b_c^\star=\infty$ since there is no switch-on 
value of $b_c$ for class $2$. Fig.~\ref{fig:2Db} represents the region where the 
switch-on of the less popular content is possible as it can be derived from the expression 
\eqref{eq:bcstar} as a function of $1/\Gamma$ and $\Lc{1}$.


\section{Game Model for Content Providers}\label{sec:game}


So far the caching rate $b_c$ has been input for the CPs in order to decide 
how to optimize the caching policy $\buc$. Let MNO propose to CPs costs 
$\lambda_c$ per caching slot. CP $c$ strategy in turn is the number $b_c$ of 
caching slots he reserves per day, with convex and compact strategy set $[0,Bc]$. 
The best response $b_c^*$ of CP $c$ depends on his contents, and his opponents' strategies. 
It is the minimizer of the {\em cost function} $U_c(b_c,b_{-c},\buc) + \lambda_c \cdot b_c$: it solves
\begin{eqnarray}\label{eq:userlutil}
&&\min_{b_c}  U_c(b_c,b_{-c},\buc) + \lambda_c \cdot b_c \\
&& \quad 0 \leq b_c \leq B_c\nonumber\nonumber
\end{eqnarray}
Here $b_{-c}=\sum_{v\not = c} b_{-v}$ and opponents' strategy profile writes $\mathbf b=(b_1,\ldots,b_{c-1},b_{c+1},\ldots,b_C)$. 

The $\buc$ appearing in \eqref{eq:userlutil} is a general caching policy 
and we shall consider two cases.
 
{\noindent \bf Caching Rate Optimizers.} In this case, the best response of content providers is 
decided for a {\em fixed caching policy} $\buc$. I.e., each content provider decides beforehand 
the caching policy $\buc$ for any given caching rate $b_c$. Let $V_c(x_c)= \sum_i g_c \, e^{-\Lc{i} x_c}$: it 
is convex and decreasing and $U_c(b_c,b_{-c},\buc)=V_c(b_c/(\sum b_c + \delta))$. Hence, if all players
 are caching rate optimizers, the game is a variant of the Kelly mechanism~\cite{YangPerfEval}. The basic 
 Kelly mechanism allocates a divisible resource among players  proportionally to the players' bids, in our 
 case the equivalent required caching rates. Here, compared to the standard formulations in 
 literature~\cite{comsnet2014,johari2004,YangPerfEval,basar} our formulation combines three specific 
 features which render it non standard: 
\begin{itemize}
\item bounded compact and convex strategy set;
\item $\delta > 0$ is equivalent to a bidding reservation, as described in \cite{basar};
\item prices may depend on the player, i.e., the game is a generalized Kelly mechanism~\cite{YangPerfEval}
\end{itemize}
We denote the Kelly mechanism in the form outlined above a {\em generalized Kelly mechanism with reservation 
and bounded strategy set}.

{\noindent \bf Simultaneous Optimizers.} In this case $\buc=\buc^*$. When players are simultaneous 
optimizers, the structure of the game still resembles the Kelly mechanism \cite{johari2004}. For 
$M=1$, the game corresponds to the case of caching rate optimizers. For $M\geq 2$, the fact 
that the game is actually a Kelly mechanism is proved formally in the following 
\begin{lma}[Kelly form for Simultaneous Optimizers]\label{lma:kellyform}
If players are simultaneous optimizers, the game \eqref{eq:userlutil} is a generalized Kelly 
mechanism with reservation and bounded strategy set.
\end{lma}
The proof of the above result is found in the Appendix. 
Here, it is sufficient to observe that even in the case of a simultaneous optimizer CP $c$, the optimal MCR can be expressed as $U_c(b_c,b_{-c})=U_c(b_c,b_{-c},\buc^*)=V_c(b_c/(\sum b_c + \delta))$ where $V_c(x_c)$ is convex and continuously differentiable in $x_c=\frac{b_c}{\sum b_c + \delta}$.

\subsection{Existence and uniqueness of the Nash Equilibrium}

In the general case, the game may comprise a mixture of both CPs who are caching rate optimizers 
and who are simultaneous optimizers. From the above discussion, the game is still a generalized Kelly 
mechanism with reservation and bounded strategy set. 

In order to characterize the possible equilibria, we describe first the best response 
$b_c^*$ of each player 
\begin{lma}[Best response]\label{lma:bestresp}
Given the opponent CPs' strategy profile $\mathbf b_{-c}$:\\
\noindent i. It holds $b_c^*=0$ if and only if $\dot U(0,b_{-c}) > -\lambda_c$ where 
\[
\dot U(0,b_{-c})=
\begin{cases}
-\frac{\sum_i \dci \Lc{i} \uci}{b_{-c}+\delta} & \mbox{caching rate optimizers} \\
-\frac{\dc{1} \Lc{1} }{b_{-c}+\delta} & \mbox{simultaneous optimizers} 
\end{cases}
\]
\noindent ii. Let $b_c^*>0$, then $b_c^*=\min\{b_c,B_c\}$, where $\dot U_c(b_c,b_{-c})=-\lambda_c$.\\
\end{lma}
The above statement follows from the fact that the objective function in \eqref{eq:userlutil} is 
convex and thus has a unique minimum in $[0,B_c]$. The expression of $\dot U(0,b_{-c})$ in the case of 
simultaneous optimizers is derived from the expression \eqref{eq:utilzero} reported in 
Lemma~\ref{lem:conv}.

The zero $\bb^*=\0$ and the saturated $\bb^*=\mathbf B$ Nash equilibria are easily characterized
 in the following 
\begin{prop}[Trivial Nash Equilibria]\label{thm:trivialNash}
\noindent i. $\bb^*=\0$ is the unique Nash equilibrium iff $\dc{1} \Lc{1} <\lambda_c \delta $ 
if $c$ is a simultaneous optimizer and $\sum \dci \Lc{i} <\lambda_c \delta $ if $c$ is a caching 
rate optimizer.\\ 
\noindent ii. $\bb^*=\mathbf B$ is the unique Nash equilibrium if and only if it holds 
$\dot U_c(B_c,\sum B_c + \delta)>-\lambda_c$ for all $c\in \C$.
\end{prop}

We observe that in the original Kelly mechanism, the strategy vector $\0$ is never a Nash 
equilibrium~\cite{comsnet2014,johari2004}. 

In our case, it may be the Nash equilibrium  and this is the effect of the term $\delta>0$ due the 
MNO's usage of the cache. In fact, the physical interpretation is provided by the condition i. in 
Prop.~\ref{thm:trivialNash}. No CP has incentive to start caching at give price 
when the marginal revenue for starting caching, i.e., represented by the product of demand and availability, does not exceed the value of the cache share reserved to the MNO operations, the term $\lambda_c \delta$. Conversely, at low prices a saturated Nash equilibrium 
$\bb^*=\mathbf B$ is expected. 

In the general case, the presence of a bounded strategy set requires a specific proof for 
the uniqueness of the Nash equilibrium, as seen in the following. %
\begin{thm}[Existence and Uniqueness]\label{lma:uniqueness}
The game has a Nash equilibrium and it is unique.
\end{thm}

We describe a brief outline of the full proof of the above result which is found in the Appendix. 
 In order to prove the existence of Nash equilibria of the game, it is sufficient
to observe that: 
\begin{itemize}
\item the multistrategy set is a convex compact subset of $\mathbb{R}^C$;
\item $U_c(b_c,b_{-c},\buc)$ is convex conditionally to the opponents strategy, both for simultaneous optimizers
and caching rate optimizers;
\end{itemize}
Hence, the existence of Nash equilibria is a direct consequence of the result of Rosen~\cite{rosen}, 
originally formulated for $n$--persons concave games. With respect to the uniqueness, $\bb^* = \0$ and $\bb^* = \mathbf B$ are always unique from Prop.~\ref{thm:trivialNash}. Once we excluded those trivial cases, the uniqueness can be derived by extending an argument~\cite{basar} to the case of a bounded strategy set. Such proof applies to both the case of simultaneous optimizers and of caching-rate optimizers. However, it requires cost functions to be continuously differentiable in $b_c$, which is not straightforward for simultaneous optimizers.

Finally, the proof of uniqueness applies also to the context where part of the players 
are simultaneous optimizers and the others are caching-rate optimizers. 

We further observe that from the proof of Thm.~\ref{lma:uniqueness} we can derive a simple bisection algorithm 
to calculate the unique solution of the game. It will be used in the numerical section where we shall provide 
further characterization of the game via quantitative measures, There, we are describing the pricing operated 
by the MNO and the convergence to the Nash equilibrium when CPs are myopic cost minimizers.


\section{Numerical Results}\label{sec:numres}


\begin{figure*}[t]
    \centering  
     		   \subfigure
    {
        \includegraphics[width=0.22\linewidth]{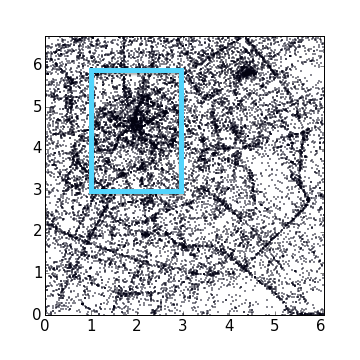}
        \put(-125,95){(a)}  
        \label{fig:F11} 
     }
  		   \subfigure
    {
        \includegraphics[width=0.23\linewidth]{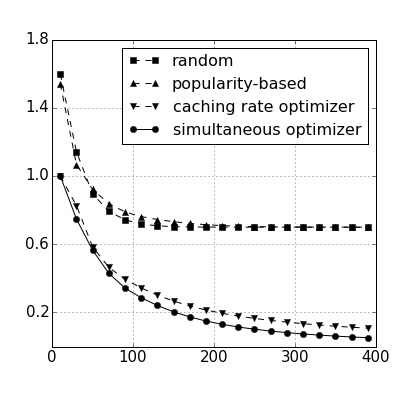}     
        \put(-129,95){(b)} 
        \put(-115,28){\begin{rotate}{90}{\small Cost Function}\end{rotate}}
        \put(-65,0){\small $r$ [m]}
        \label{fig:F12} 
     }
          \subfigure
    {
        \includegraphics[width=0.23\linewidth]{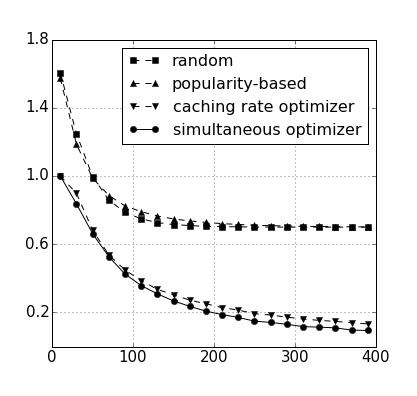}
        \put(-125,95){(c)}         
        \put(-115,28){\begin{rotate}{90}{\small Cost Function}\end{rotate}}
        \put(-65,0){\small $r$  [m]}
        \label{fig:F13}
     }
      \subfigure
    {
        \includegraphics[width=0.23\linewidth]{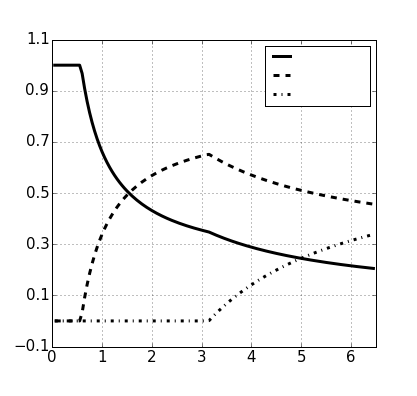}
        \put(-125,95){(d)} 
        \put(-125,50){${\uci}^*$}
        \put(-65,0){\small $b_c$}
        \put(-32,95){\scriptsize $i=1$} 
        \put(-32,89){\scriptsize $i=2$} 
        \put(-32,83){\scriptsize $i=3$} 
        \label{fig:F14}
     }
       \caption{(a) Milan downtown base stations deployment, detail of the area considered; (b) CP optimal cost: theoretical prediction (c) CP optimal cost: outcome of the simulation (d) CP optimal caching policy for varying $b_c$ and fixed value of $r=210$ m. Settings are: $M=3$, $\dci= 0.589$,  $0.294$,  $0.118$, $\delta= 2$, $b_c=70$, $b_{-c}=300$, $N=10000$, $\nci=1000$, $4000$, $10000$.}\label{fig1}
\end{figure*}
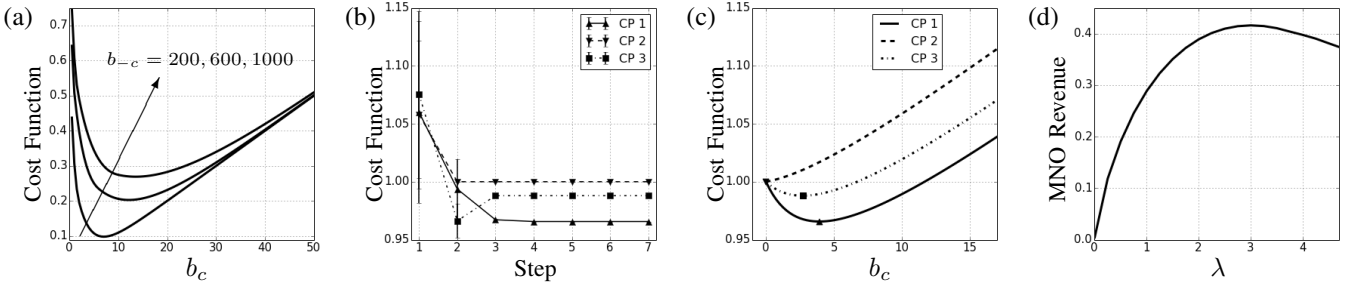
\begin{figure*}[t]
    \centering
       	   \subfigure 
    {
        \includegraphics[width=0.23\linewidth]{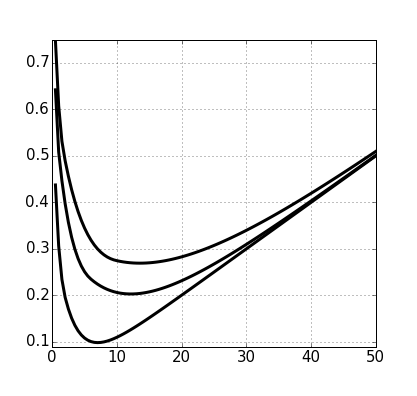}
        \put(-129,95){(a)}
        \put(-90,80){\scriptsize $b_{-c}=200,600,1000$}\put(-100,15){\vector(1,2){30}}
        \put(-115,28){\begin{rotate}{90}{\small Cost Function}\end{rotate}}
        \put(-60,0){$b_c$}
        \label{fig:F21}
     }    
   	   \subfigure 
    {
        \includegraphics[width=0.23\linewidth]{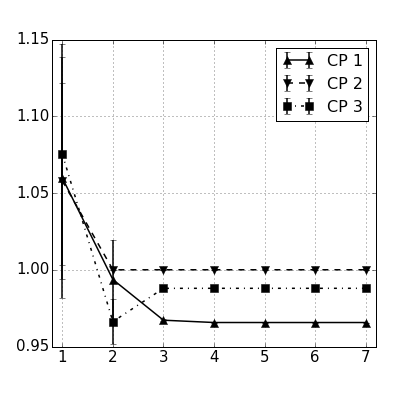}
        \put(-129,95){(b)} 
        \put(-115,28){\begin{rotate}{90}{\small Cost Function}\end{rotate}}
        \put(-65,0){\small Step}
        \label{fig:F22}
    }     
  		   \subfigure
    {
        \includegraphics[width=0.23\linewidth]{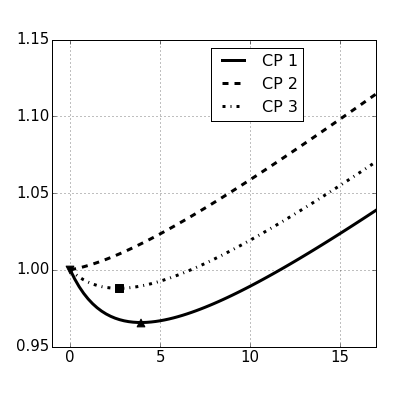}
        \put(-129,95){(c)}
        \put(-115,28){\begin{rotate}{90}{\small Cost Function}\end{rotate}} 
        \put(-60,0){$b_c$}
        \label{fig:F23} 
     }
       \subfigure
       {
        \includegraphics[width=0.23\linewidth]{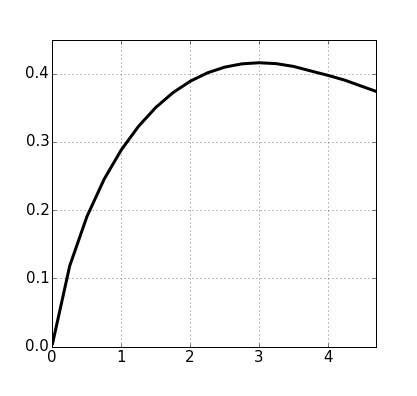}
        \put(-129,95){(d)}
        \put(-115,28){\begin{rotate}{90}{\small MNO Revenue}\end{rotate}}
        \put(-60,0){$\lambda$}
        \label{fig:F24} 
     }
   \caption{(a) Cost function for simulataneous optimizer CP as $b_c$ varys, all parameters are the same as in Fig~\ref{fig:F14}; (b) Dynamic of the cost function for a 3-simoultaneous optimizers game. Settings are: $g_1=[0.18,0.27,0.55]$, $g_2=[0.3,0.6,0.1]$, $g_3=[0.6,0.1,0.3]$, $\Lambda_1^1=\Lambda_1^2=\Lambda_1^3=0.3$, $\Lambda_2^1=\Lambda_2^2=\Lambda_2^3=0.1$, $\Lambda_3^1=\Lambda_3^2=\Lambda_3^3=0.2$  (c) detail of the corresponding restpoint; (d) revenue of the MNO for increasing uniform price. Parameters are: $g_1=[0.3,0.2,0.5]$, $g_2=[0.3,0.5,0.2]$, $g_3=[0.29,0.36,0.35]$, $N=70$, $N_1^i=600$, $N_2^i=700$, $N_3^i=500$, $\delta=2$, $r=73$m.} \label{fig2}
\end{figure*}

In this section we provide numerical description and validation of the model.\footnote{Both the Python 
scripts and the dataset used for validation can be downloaded at \url{https://www.dropbox.com/s/mm1hja2dbp4tw0x/caching_scripts.tar.gz?dl=0}.} 
First we validate the models' assumptions against a real world scenario. Then, we focus on the single player's actions, having fixed the remaining players' strategies. Finally we provide numerical characterization of the game introduced in the 
previous section.\\
{\noindent \em Point Process.} The model introduced in Sec.~\ref{sec:model} assumes that SCs are 
distributed according to a spatial Poisson process of given intensity $\Lambda$. Hence, we have tested  
the performance of the optimal caching policy in the case the SCs spatial deployment does not adhere 
to the assumption of a Poisson point distribution. In order to do so, we have been comparing the 
theoretical results with the outcome of a simulation performed over a real dataset. The real dataset 
(source \url{http://opencellid.org/}) is the sample distribution of the cell towers deployed in 
downtown Milan over a $ 2\times 3$ Kms area, as depicted in Fig.~\ref{fig:F11}: it includes the 
location of $4717$ cell towers corresponding to $\Lambda=786.2$ base stations per square Km. The 
distribution of base stations in a very densely populated urban area has been used as a reasonable 
approximation for a SC deployment. 

The sample spatial density  $\Lambda$ has been used in the model in order to evaluate, under 
the same spatial density of SCs, the theoretical CP's cost function for increasing values of the covering radius 
$0\leq r\leq 400$m in the following cases (see Fig.~\ref{fig:F12}): a) the CP performs a uniformly {\em random} 
caching policy $\uc{i}=1/3$, $i=1,2,3$ for constant caching rate $b_c$ b) the CP performs a {\em popularity-based} 
caching policy, i.e., $\uc{i}:=\dci/\sum \dci$, for constant $b_c$  c) the CP is a {\em caching rate optimizer} 
adopting a popularity based caching policy d) the CP is a {\em simultaneous optimizer}. 

The results in Fig.~\ref{fig:F13} refer to a simulation encompassing the same strategies under the 
sample point distribution of  Fig.~\ref{fig:F11}. The simulation has been performed by repeatedly selecting 
a random UE position in the playground, and measuring the sampling frequency of missed cache events upon
 requesting contents from SCs within the UE's radio range. 

By comparing the results in  Fig.~\ref{fig:F12} and  Fig.~\ref{fig:F13}, we observe that the 
Poisson distribution -- as expected due to the non-uniform spatial density of the sample real-world 
deployment -- tends to slightly underestimate the cost incurred by CPs. However, the theoretical 
and the simulated results are very close and the relative performance of the caching policies match 
the prediction of the theoretical model. This result confirms that the proposed model performs 
well even in real world scenarios: under a non-Poisson point process for the SC spatial distribution 
a rational optimizing player would choose the proposed optimal strategy over other possible strategies.

{\noindent \em Cost function.} In the next experiment we describe the optimal caching policy 
(Fig.~\ref{fig:F14}) and the cost function (Fig.~\ref{fig:F21}) in the case $M=3$. In particular, 
Fig.~\ref{fig:F14} reports on the characteristic waterfilling structure of the optimal caching as 
the parameter $b_c$ increases. As predicted by the model, the water-filling solution has a threshold 
structure. The value of $b_c$ determines the content classes that become active: for large $b_c$ all 
content classes are cached, whereas for small values only some are cached. In Fig.~\ref{fig:F21} we 
have reported the typical convex shape of the cost function corresponding to the same setting and for 
increasing values of $b_{-c}$. It is worth noting how the actions of opponents, reflected in the value 
of $b_{-c}$, affect the shape of $c$'s cost function.

{\noindent \em Convergence to the Nash equilibrium.} In Fig.~\ref{fig:F22} and Fig.~\ref{fig:F23} 
we have simulated $3$ CPs who are simultaneous optimizers. They behave as {\em myopic} players: each one 
of them, chosen at random, optimizes his own cost function based on the opponents' profile. Numerical 
simulations show that, after a small number of iterations the game stabilizes on the same restpoint 
irrespective of initial strategies. As depicted in Fig.~\ref{fig:F23} the restpoint is indeed a 
minimum for each CP's cost function, i.e., it is the Nash equilibrium of the game. This behavior 
suggests that the game has the finite improvement property~\cite{MondreShapPotGames}, even though we could 
not identify analytically a potential for the game. Hence, the system would naturally converge 
to his unique Nash equilibrium if each player optimizes independently its own cost function against 
the opponents. 

Finally, we have drawn in Fig~\ref{fig:F24} the daily revenue of the MNO
at the Nash equilibrium $\bb^*$ as a function of the caching price $\lambda$, uniform for all CPs. Because the MNO's total revenue $\sum_c \lambda \cdot b_c^*$ depends on the Nash equilibrium, she could try to optimize her revenue by leveraging the CPs' cost structure. We observe  numerically that the total revenue appears to have a unique maximum at a certain maximizer price $\lambda^*$. This suggests the existence of a unique Stackelberg  equilibrium for the proposed scheme. This provides the possibility to compute the global restpoint of the
 system when both CPs and MNO behave strategically.


\section{Conclusions}\label{sec:concls}


A model for mobile edge caching in 5G networks has been presented. OTT 
content providers compete for the cache memory made available by a MNO 
at given price. Several features of the system are captured, including 
popularity and availability of contents, spatial distribution of small cells, 
competition for cache memory and the effect of price. CPs can optimize the 
allocation of contents in order to reduce customers' aggregated missed 
cache rate. We have found that the optimal caching policy is of waterfilling 
type. Also, it is showed to give priority to contents based on popularity 
and availability. We have confirmed the validity of the caching policy 
optimization on real-world traces. 
 
Finally, the competition for the shared caching memory can be formulated
as a convex $n$--persons game. This game is a new form of the Kelly mechanism 
with bounded strategy set, where each CP trades off the expected missed cache 
rate for the price paid to the MNO in order to reserve cache memory space. The 
existence and uniqueness properties of the Nash equilibrium  
are demonstrated. Also, numerical results indicate that when CPs are 
myopic optimizers, the system converges to a unique restpoint which is the 
Nash equilibrium. 

Furthermore, from numerical results, this game appears to have a unique Stackelberg 
equilibrium, a relevant feature for the MNO in order to maximize her revenue at
 the optimal price. To this respect, an interesting research direction is to develop 
 online algorithms by which the MNO can learn over time such optimal price.

\bibliographystyle{IEEEtran}
\bibliography{CachingBib}  


\section{Appendix}


\subsection{Proof of Lemma~\ref{lemma:saturation}}

\begin{IEEEproof}
Let ${\mathbf u}_c^*$ the optimal allocation, and let us assume that ${u^*}_c^{r+1}>0$
It is sufficient to write the generic cost as 
\[
U_c(b_c,b_{-c},\buc)=\sum_{i \in I} \dci \, e^{-\pi r^2\Lambda \frac{N}{N_i}\frac{\uci b_c}{b+\delta}}+\sum_{i \not \in I} \dci
\]
from which it is immediate to see that a response ${\mathbf u}_c$ identical to ${\mathbf u}_c^*$ but where 
$\uc{r}={u^*}_c^{r+1}$ and $\uc{r+1}={u^*}_c^{r}$ is better off. In fact we can write
\begin{eqnarray}
&& \hskip-7mm \Delta U=  U_c(b_c,b_{-c},\buc^*) - U_c(b_c,b_{-c},\buc) \nonumber \\
&&\hskip-7mm =\dc{r+1} e^{-\pi r^2\Lambda \frac{N}{\nc{r+1}}\frac{{\uc{r+1}}^* b_c}{b+\delta}} \hskip-2mm + \dc{r} - \dc{r} e^{-\pi r^2\Lambda \frac{N}{\nc{r}}\frac{{\uc{r+1}}^* b_c}{b+\delta}}-\dc{r+1}\nonumber \\
&& \hskip-7mm > \dc{r+1} e^{-\pi r^2\Lambda \frac{N}{\nc{r}}\frac{{\uc{r}}^* b_c}{b+\delta}} + \dc{r} - \dc{r} e^{-\pi r^2\Lambda \frac{N}{\nc{r}}\frac{{\uc{r}}^* b_c}{b+\delta}}-\dc{r+1}\nonumber  \nonumber \\
&& \hskip-7mm > (\dc{r} - \dc{r+1})\Big ( 1 - e^{-\pi r^2\Lambda \frac{N}{\nc{r}}\frac{{\uc{r}}^* b_c}{b+\delta}} \Big ) > 0 \nonumber
\end{eqnarray}
which concludes the proof.
\end{IEEEproof}

\subsection{Proof of Lemma~\ref{eq:utilzero}}
\begin{IEEEproof}
One only content type is cached if and only if
\begin{eqnarray}
&&1=\uc{[1]} =\frac{b_c+u_{-c}+\delta}{\Lc{[1]}}\left(\log(1/\nu) -\log(\alpha_{[1]})\right)\nonumber\\
&&\log(1/\nu) -\log(\alpha_{[i]})<0, \forall i\geq 1
\end{eqnarray}
After simple calculations, the above condition brings
\begin{eqnarray}
&&1/\nu=e^{\frac{\Lc{[1]}b_c}{b_c+u_{-c}+\delta}}\alpha_{[1]}< \alpha_{[i]}, \forall i>1\nonumber
\end{eqnarray}
Which can hold true if and only if the exponential term is close enough to $1$. Finally, observe 
that the ordering of the $\alpha_i$'s does not depend on $b_c$ and that the exponential converges to 1 for small $b_c$, which completes the proof.
\end{IEEEproof}

\subsection{Proof of Lemma~\ref{lemma:lemma3}}
\begin{IEEEproof}
If we assume $H$ and $f$ are both twice differentiable the proof
is trivial. 

In the case of non differentiable functions, we can verify the 
convexity condition for $f(\cdot)$. In fact, for 
every $0\leq x_1 \leq x_2$ and $t\in (0,1)$: 
\begin{eqnarray}\label{eq:conv}
&&f(tx_1+(1-t)x_2)=\frac{H(tx_1+(1-t)x_2)}{tx_1+(1-t)x_2}\\
&&                \leq \frac{tH(x_1)+(1-t)H(x_2)}{tx_1+(1-t)x_2}=g(t)
\end{eqnarray}
Now, we observe that $g(0)=f(x_2)$ and $g(1)=f(x_1)$. Also, we note that 
$g$ is convex for $t\in [0,1]$. In fact, $g$ is differentiable in $t$, 
and, by rewriting $g(t)=N(t)/D(t)$, a direct calculation provides
\[
\ddot g(t)=\frac{2x_1x_2(x_2-x_1)\Big (\frac{H(x_1)}{x_1}-\frac{H(x_2)}{x_2} \Big )}{(t\,x_1+(1-t)\,x_2)^3}\geq 0
\]
where the nonnegative sign is due to assumption that $f$ is non-increasing. Now, by convexity, we can write  
\[
g(t)=g((1-t)\cdot 0 +t\cdot1)\leq (1-t)g(0)+tg(1)
\]
and replacing the above expression in \eqref{eq:conv}, $f$ is seen to satisfy 
the claim of convexity.
\end{IEEEproof}

\subsection{Proof of Lemma~\ref{lma:kellyform}}
\begin{IEEEproof}
First, we have to prove that the optimal cost can be expressed as $U_c(b_c,b_{-c})=V_c(b_c/(\sum b_c + \delta))$ where $V_c(x_c)$ is convex and continuously differentiable in $x_c=\frac{b_c}{\sum b_c + \delta}$. In order to do so, 
 for a given value of $b_{-c}$ we denote $\mathcal B(b_{-c})=\{b_{[1]}^\star,\ldots,b_{[M]}^\star\}$ the set of thresholds such that, for $b_{[k]}^\star\leq b_c <b_{[k+1]}^\star$ it holds $\uc{[1]},\ldots,\uc{[k]}>0$ and $\uc{[k+1]}=\ldots=\uc{[M]}=0$. Hereafter, let us simplify the notation and consider the indexes sorted according to $[\cdot]$ as described in Sec.~\ref{sec:contalloc}. %
From \eqref{eq:watersol1} and accounting for the expression of the $\alpha$s, it is immediate to see that the number  of active content classes is a function of the type $k=k\Big( \frac{b_c}{\sum b_c + \delta}\Big)$. Also, for $b^\star(k)\leq b_c< b^\star(k+1)$, we can calculate the closed form 

\begin{eqnarray}\label{eq:optcostexpl}
U(b_c,b_{-c})=B_k \prod_{i=1}^k (\Lci \dci)^{\frac{1}{B_k\Lci}} e^{-\frac{1}{B_k}\frac{b_c}{b_c+b_{-c}+\delta}}+%
\sum_{i=k+1}^M \dci
\end{eqnarray}

where $B_k=\sum_{i=1}^k \frac1{\Lci}$. Hence, by inspection, the right hand term is a function of 
$x_c=\frac{b_c}{\sum b_c + \delta}$. 

Now, we have to prove that such function if continuously differentiable. Since $U(b_c,b_{-c})$ is smooth in $(b_c^\star(k),b_c^\star(k+1))$, we can restrict to the threshold points, in particular, we consider $b_{k+1}^\star$ and verify that equality holds for the left and right derivative of \eqref{eq:optcostexpl}. It holds
\begin{equation}\label{eq:derivative}
\frac{d}{db_c} U(b_c,b_{-c})=-\prod_{i=1}^k (\Lci \dci)^{\frac{1}{B_k\Lci}} e^{-\frac{1}{B_k}\frac{b_c}{b_c+b_{-c}+\delta}}\frac{d}{db_c} \frac{b_c}{\sum_v b_v+\delta} 
\end{equation}
The value of threshold $b^\star=b^\star(k+1)$ is derived by the relation $\log(1/\nu)=\log(\alpha_{k+1})$ which writes 
\begin{small}
\begin{eqnarray}\label{eq:bstarval}
\frac{b^\star}{\sum b^\star + b_{-c} + \delta}=-\sum_{i=1}^k\frac{\log(\alpha_i)}{\Lci}+A_k\log \Big ( \frac{\sum b^\star + b_{-c} + \delta}{\Lc{k+1} \dc{k+1} b^\star }\Big )
\end{eqnarray}
\end{small}
We can now replace \eqref{eq:bstarval} and  $\alpha_{i}=\frac{b^\star+b_{-c}+\delta}{\dci \Lci b^\star}$  into \eqref{eq:derivative} for both the case $k$ and $k+1$. A direct calculation, which we omit for the sake of space, shows that the equality at $b^\star$ is verified, concluding the statement.  
\end{IEEEproof}

\subsection{Proof of Theorem~\ref{lma:uniqueness}}

The proof of Theorem \ref{lma:uniqueness} will require the result 
in the Lemma reported next. The Lemma itself is a technical continuity argument. 
We remark that, even in the unbounded case, the results of uniqueness of the 
Nash equilibrium for the Kelly mechanism \cite{basar}, requires the cost function 
to be twice continuously differentiable.

\begin{lma}\label{lemma:xFunction}
For each player $c$ there exists a unique, continuous, decreasing function $x_c(p)$ such that 
\begin{equation}\label{eq:implicit}
\frac{d}{d x}V_c(x_c(p))(1-x_c(p))+p\lambda_c=0, 
\end{equation}
where $V_c(x)$ is the function defined in the proof of Lemma \ref{lma:kellyform}.
\end{lma}

\begin{IEEEproof}
Since we refer to player $c$, unless required for the sake of clearness, we shall hereafter 
identify $x:=x_c$ for the sake of notation. First, we need to have a closer look to the structure of the cost function. We know that there exist $n$, $n\leq M-1$, intervals for $b_c$ such that in each interval the cost function is described as in (\ref{eq:optcostexpl}). Now, let $x=x(b_c,b_{-c})=\frac{b_c}{b_c+b_{-c}+\delta}$. We want to characterize $x_k^\star=\frac{b^\star_k}{b^\star_k+b_{-c}+\delta}$. By definition, $x^\star_k$ is the threshold value above which the $k$-th content type will start to be cached. 

Now, resorting to the water-filling formulation of the optimal allocation in (\ref{eq:watersol1}), the $k$-th threshold is uniquely identified by the condition 
\begin{equation}\label{eq:condthrak}
\begin{cases}
1/\nu=\alpha_k \\
\sum^{k-1}\frac{1}{x^\star_k\Lambda_c^i}(\log(\frac{1}{x^\star_k\Lambda_c^kg_c^k})-\log(\frac{1}{x^\star_k\Lambda_c^ig_c^i}))=1
\end{cases}
\end{equation}
The condition \eqref{eq:condthrak}

\begin{equation}\label{eq:criticx}
x^\star_k=\sum^{k-1}\frac{1}{\Lambda_c^i}\log \left (\frac{\Lambda_c^i g_c^i}{\Lambda_c^k g_c^k} \right )
\end{equation}

It is important to observe that the $x^\star_k$s do not depend on $b_{-c}$: the $M$ critical values for $x$ are the values $x^\star_k$  determined by \eqref{eq:criticx} such that $x^\star_k\leq 1$. At this point, we observe that the thresholds in $x$ for player $c$ depend on $c$'s parameters only, and they are naturally ordered increasing with the lag $k$ and. Incidentally, we observe that, since $x\in[0,1]$, it is well possible that 
just some of the $x^\star_k$ are smaller than one, and those that are bigger than one are not attained. In fact, 
this means that some content classes may not be cached, for any value of $b_c$.

We can now go on with the main proof of the Lemma. In \cite{basar} the cost function is assumed to be twice continuously differentiable, i.e., $C^2$. However, our cost function is only continuously differentiable in the variable $b_c$, i.e., $C^1$. More in detail, by direct inspection it is possible to verify that that the optimal missed cache rate is $C^2$ piecewise, with a finite number of points where the second order derivative is discontinuous: those correspond precisely the values $x_{k}^{\star}$ discussed above. 

The argument proceeds as follows. Let $U_c(b_c,b_{-c}) +\lambda_c(b_c)$ be the cost function of a simultaneous optimizer. The best response of the player $c$ must fulfill the following relation: 
\begin{equation}\label{eq:dU}
\frac{d}{d b_c}U(b_c,b_{-c})+\lambda_c=0.
\end{equation}
 Now, define 

 $$V(x)=B_k \prod_{i=1}^k (\Lci \dci)^{\frac{1}{B_k\Lci}} e^{-\frac{1}{B_k}x}+\sum_{i=k+1}^M \dci,\ x\in]x^\star_k,x^\star_{k+1}[.$$
We know that $V\in C^2([0,1]\setminus\{x^\star_1,...,x^\star_M\})\cap C^1([0,1])$. Letting $x=\frac{b_c}{b_c+b_{-c}+\delta}, p=b_c+b_{-c}+\delta$, we can rephrase (\ref{eq:dU}) as
\begin{equation}\label{eq:dUx}
\frac{d}{d x}V(x)(1-x)+p\lambda_c=0,\ x\in]x^\star_k,x^\star_{k+1}[.
\end{equation}
Now, in $]x^\star_k,x^\star_{k+1}[\times \mathbb{R}$ the left hand side is a $C^1$ function (indeed it is $C^\infty$, but $C^1$ is sufficient for our argument), hence we can apply the implicit function theorem and, in the same fashion as in \cite{basar}, we can derive the existence of a continuous decreasing function $x:]p_k,p_{k+1}[\rightarrow]x_k,x_{k+1}[$, such that $\frac{d}{d x}V(x(p))(1-x(p))+p\lambda_c=0,\ p\in]p_k,p_{k+1}[$. Moreover, the monotonicity of $x$ implies 
\[
\lim_{p\rightarrow p_{k}}x(p)=x^\star_{k}, \lim_{p\rightarrow p_{k+1}}x(p)=x^\star_{k+1}
\]

We repeat the same argument on each interval $]x^\star_k,x^\star_{k+1}[$, and we finally glue together the $x$ functions defined over each interval obtaining a monotone surjective function. From the monotonicity of $V$, such  function is well posed since \eqref{eq:dUx} forces same values of $x$ for same values of $p$. Also, continuity and monotonicity ensure that the the domain of $x$ is a connected set, i.e., the interval $\cup ]p_k,p_{k+1}[=(0,\delta+\sum B_c)$. 

Thus, we have defined a unique decreasing, continuous function 
$x(p):[0,\sum B_c + \delta]\rightarrow[0,1]$ such that $(x(p),p)$ solves (\ref{eq:dUx}). 
\end{IEEEproof}
\hspace{2mm}

We can now prove Theorem~\ref{lma:uniqueness}.\\[2mm]

\begin{IEEEproof}
In order to characterize the existence of Nash equilibria of the game, it is sufficient to observe that: 
\begin{itemize}
\item the strategy set is a convex compact subset of $\mathbb{R}^C$;
\item $U_c(b_c,b_{-c},\buc)$ is convex conditionally to the opponents strategy;
\end{itemize}
Hence, the existence of Nash equilibria is a  direct consequence of the result of Rosen~\cite{rosen}, 
originally formulated for $n$--persons concave games (here players minimize so convexity applies). 
Let $\bb^*$ is a Nash equilibrium:  $\bb^* = \0$ and $\bb^* = \mathbf B$ are always unique from 
Prop.~\ref{thm:trivialNash}. Let us hence consider the remaining possible equilibria.

Let $x_c(p)$ the function defined at (\ref{eq:implicit}). From Lemma \ref{lma:bestresp}, we can 
now write the best response in $x$ for each player as a function of $p$:
\begin{equation}\label{eq:demand}
\hat x_c(p)=\max(0,\min(x_c(p),B_c/p)).
\end{equation}
Since all functions in the definition of $\hat x_c(p)$ are decreasing and continuous in $p$, so it is $\hat x_c(p)$. Moreover, $\hat x_c(0)=1, \lim_{p\rightarrow\infty}\hat x_c(p)=0$.
Now, we observe that the actual best responses of players in a Nash 
equilibrium need to satisfy the condition

\begin{equation}\label{eq:soldemand}
\sum \hat x_c(p) = 1 - \frac{\delta}{p},\  p\in\left[0,\sum B_c+\delta\right].
\end{equation}
But condition \eqref{eq:soldemand} determines a unique $p$. In fact observe that the sum on the left-hand side is decreasing in $p$, its value at $0$ is $M$ and tends to $0$ as $p$ increases. The term on the right-hand side is increasing in $p$ and tends to 1 as $p$ increases. It follows that the two functions can be equal in no more than one point. This implies that the Nash equilibrium is unique and is determined by the unique $p^\star$ such that the equality holds true.

We can define $C_0(p)=\{c \in \C| \hat x_c(p)=0\}$. Also, the set 
$\C_B(p)=\{c \in \C| p \cdot \hat x_c(p) \geq  B_c\}$ is unique for every value of 
$p \in [0,\sum B_c)$. 
Finally, the Nash equilibrium $\bb^*$ is derived by the bijection $\bb=\phi(p^*)$, where\\
\noindent i. $\phi_c(p^*) = 0$ for $c\in \C_0(p^*)$;\\
\noindent ii. $\phi_c(p^*) = B_c$ for $c\in \C_B(p^*)$;\\
\noindent iii. the $\phi_c(p^*)$s for $c\in  \C'(p^*)=\C \setminus (\C_0(p^*)\cup\C_B(p^*))$ with the bijection 
induced from the full rank compatible linear system
\[
b_c^*(1-x_c^*)+\sum_{v\in \C'(p^*)} b_v^* x_v^*  =-\delta x_c^* - |\C'(p^*)| \frac{B^2}{p} , c\in \C'(p^*)
\]
which concludes the proof. 
\end{IEEEproof} 

\begin{remark}
It is worth observing that we proved that uniqueness holds despite the optimal cost function is not $C^2$ as required in \cite{basar}. In order to extend the argument of \cite{basar} to our case, we had to carefully adapt the implicit function theorem to the case of a piecewise-$C^1$ function. The uniqueness follows by  continuity and  monotonicity of the implicit function. 
\end{remark}

\end{document}